\documentclass[twocolumn,showpacs,preprintnumbers,amsmath,amssymb,prl,superscriptaddress]{revtex4-1}
\usepackage{bbm}
\usepackage{mathrsfs}
\usepackage{graphicx}
\usepackage{dcolumn}
\usepackage{bm}
\usepackage{amsmath}
\usepackage{amsfonts}
\usepackage{color}
\usepackage[colorlinks=true,citecolor=blue,anchorcolor=blue]{hyperref}
\usepackage{floatrow}

\begin{document}

\title{Anisotropic topological magnetoelectric effect in axion insulators}
\author{Zhaochen Liu}
\affiliation{State Key Laboratory of Surface Physics and Department of Physics, Fudan University, Shanghai 200433, China}
\author{Jing Wang}
\thanks{wjingphys@fudan.edu.cn}
\affiliation{State Key Laboratory of Surface Physics and Department of Physics, Fudan University, Shanghai 200433, China}
\affiliation{Institute for Nanoelectronic Devices and Quantum Computing, Fudan University, Shanghai 200433, China}

\begin{abstract}
Three-dimensional topological insulators or axion insulators exhibit the topological magnetoelectric effect, which is isotropic with a universal coefficient of proportionality quantized in units of $e^2/2h$. Here we study the finite-size effect of topological magnetoelectric effect, and find the magnetoelectric coefficients are anisotropic, namely $\alpha_{xx}\neq\alpha_{zz}$. Both of them are shown to converge to a quantized value when the thickness of topological insulator film $d$ increases reaching the three-dimensional bulk limit. The nonzero value of $(\alpha_{xx}-\alpha_{zz})\propto1/d$ could be measured by using the gyrotropic or nonreciprocal birefringence of terahertz light. The unique $1/d$ dependence on film thickness of the rotation angle of optical principle axes is the manifestation of topological magnetoelectric effect, which may also serve as a smoking gun signature for axion insulators.
\end{abstract}

\date{\today}


\maketitle

The search for topological quantization phenomena has become one of the important goals in condensed matter physics~\cite{thouless1998}. Two well-known examples are the flux quantization in units of $h/2e$ in superconductors~\cite{byers1961} and the Hall conductance quantization in units of $e^2/h$ in the quantum Hall effect (QHE)~\cite{thouless1982}. The exact quantization of the topological phenomena provides the precise metrological definition of fundamental physical constants~\cite{klitzing2019}.

A new topological phenomena called quantized topological magnetoelectric (TME) effect has been predicted to exist in the three-dimensional (3D) time-reversal ($
\mathcal{T}$) invariant topological insulator (TI)~\cite{qi2008,hasan2010,qi2011}, where a quantized polarization is induced by a magnetic field, and its dual, a quantized magnetization in response to an electric field. Such an electromagnetic response can be described by the \emph{rotationally invariant} topological $\theta$ term~\cite{qi2008}
\begin{equation}\label{theta_term}
\mathcal{L}_{\theta}=\frac{\theta}{2\pi}\frac{e^2}{h}\mathbf{E}\cdot\mathbf{B},
\end{equation}
together with the ordinary Maxwell Lagrangian. Here $\mathbf{E}$ and $\mathbf{B}$ are the conventional electromagnetic fields inside the insulator, $e$ is the charge of an electron, $h$ is Plank's constant, $\theta$ is the dimensionless pseudoscalar parameter known as the axion angle in particle physics~\cite{peccei1977,wilczek1987}. The quantization of $\theta=\pi$ (defined module $2\pi$) in TIs depends only on $\mathcal{T}$-symmetry and bulk topology reflecting the $Z_2$ topological index,  which is therefore universal and independent of any material details. Microscopically, $\theta$ represents the contribution to ME polarizability from extended orbitals~\cite{essin2009,coh2011}. From the effective action with an open boundary condition, $\theta=\pi$ describes a surface QHE with a half-quantized Hall conductance, which is the physical origin of TME and leads to a variety of exotic phenomena such as quantized anomalous Hall (QAH) effect~\cite{chang2013b} and topological magneto-optical effect~\cite{okada2016,wul2016,dziom2017}. However, for a finite $\mathcal{T}$-invariant TI, $\mathcal{T}$ forces TME effect to vanish, where the surface and bulk states contributions to TME effect precisely cancel each other~\cite{mulligan2013,witten2016,rosenow2017}. To observe quantized TME effect in TIs, one must fulfill three stringent requirements~\cite{qi2008,wang2015b}. First, a surface gap is induced by a hedgehog magnetization~\cite{qi2008,wang2015b,morimoto2015,mogi2017,nomura2011,tokura2019}, thus the $\theta$ value is uniquely defined, such a state is defined as axion insulator (AI)~\cite{wang2015b,morimoto2015,mogi2017,nomura2011,tokura2019,li2010,wang2017c,mogi2017a,grauer2017,xiao2018,varnava2018,wieder2018,allen2019,wan2012,turner2012}, which can be viewed as a higher-order TI with its order higher than its dimension (without any gapless surface, hinge or corner states)~\cite{varnava2018,wieder2018}. Second, the Fermi level is finely tuned into the magnetically induced surface  gap while keeping the bulk truly insulating. Third, the finite-size effect is eliminated by thick enough TI film to guanrantee the exact quantization.

TME is a hallmark of 3D TI. Several other theoretical proposals have been made to realize the TME~\cite{wang2015b,morimoto2015,qi2009b,maciejko2010,tse2010,bermudez2010,yu2019}. However, it has not yet been observed experimentally. Generically, a linear ME coupling of the following form appears in a material when both $\mathcal{T}$ and spatial inversion symmetry $\mathcal{P}$ are broken~\cite{fiebig2005},
\begin{equation}
\mathcal{L}_{\text{ME}}=\alpha_{ij}E_iB_j,
\end{equation}
$\alpha_{ij}$ is the ME susceptibility tensor, $i,j=x,y,z$. The $\mathcal{T}$-symmetry restricts the off-diagonal elements of $\alpha_{ij}$ to vanish, and the ambiguity in defining the bulk polarization~\cite{vanderbilt1993,ortiz1994} allows the diagonal elements to take a nonzero value. $\alpha_{ij}=(\theta/2\pi)(e^2/h)\delta_{ij}$ with $\theta=\pi$ in a 3D AI. Isotropiness and quantization are the characteristics of TME, which only exists in 3D bulk limit. To avoid confusion, here we are interested only in the orbital ME polarizability with topological character in $\alpha_{ij}$. For a finite AI film, TME is not quantized and the topological $\alpha_{ii}$ may be anisotropic.

In this paper, we study the finite-size effect of TME, and find TME coefficients are anisotropic (namely $\alpha_{xx}\neq\alpha_{zz}$) in a finite AI system. Both $\alpha_{xx}$ and $\alpha_{zz}$ are shown to converge to a quantized value when the thickness of AI film $d$ increases. The nonzero value of $(\alpha_{xx}-\alpha_{zz})\propto1/d$ could be measured by using the gyrotropic birefringence (GB) of terahertz light~\cite{kurumaji2017,hornreich1968} with a measurable rotation angle accessible by the current technique, and the \emph{unique} $1/d$ dependence on film thickness is the manifestation of TME effect.

\emph{Model system.} The general theory for the finite-size effect of TME is generic for any TI or AI materials. We would like to start with the newly discovered antiferromagnetic (AFM) AI MnBi$_2$Te$_4$ for concreteness~\cite{zhang2019,li2019,otrokov2019,gong2019,deng2020,liu2020}. The material consists of Van der Waals coupled septuple layers (SL) and develops $A$-type AFM order with an out-of-plane easy axis, which is ferromagnetic (FM) within each SL but AFM between adjacent SL along $z$ axis. The bulk MnBi$_2$Te$_4$ breaks $\mathcal{T}$, but $\theta=\pi$ is protected by a combined symmetry $\mathcal{S}\equiv\mathcal{T}\tau_{1/2}$, where $\tau_{1/2}$ is the half translation operator along $z$ axis. The odd SL MnBi$_2$Te$_4$ film has uncompensated FM layer and is a QAH insulator~\cite{deng2020}, while even SL film breaks both $\mathcal{T}$ and $\mathcal{P}$ but conserves $\mathcal{PT}$ and is an AI. Such an AI state is characterized by a zero Hall plateau with zero longitudinal conductance~\cite{wang2015b,wang2014a,deng2020,liu2020}. The $\mathcal{S}$-breaking surfaces are gapped by the intrinsic magnetism in this system, allowing a finite TME response. In the following, we study the TME effect in even SL film. 

\begin{figure}[t]
\begin{center}
\includegraphics[width=3.4in,clip=true]{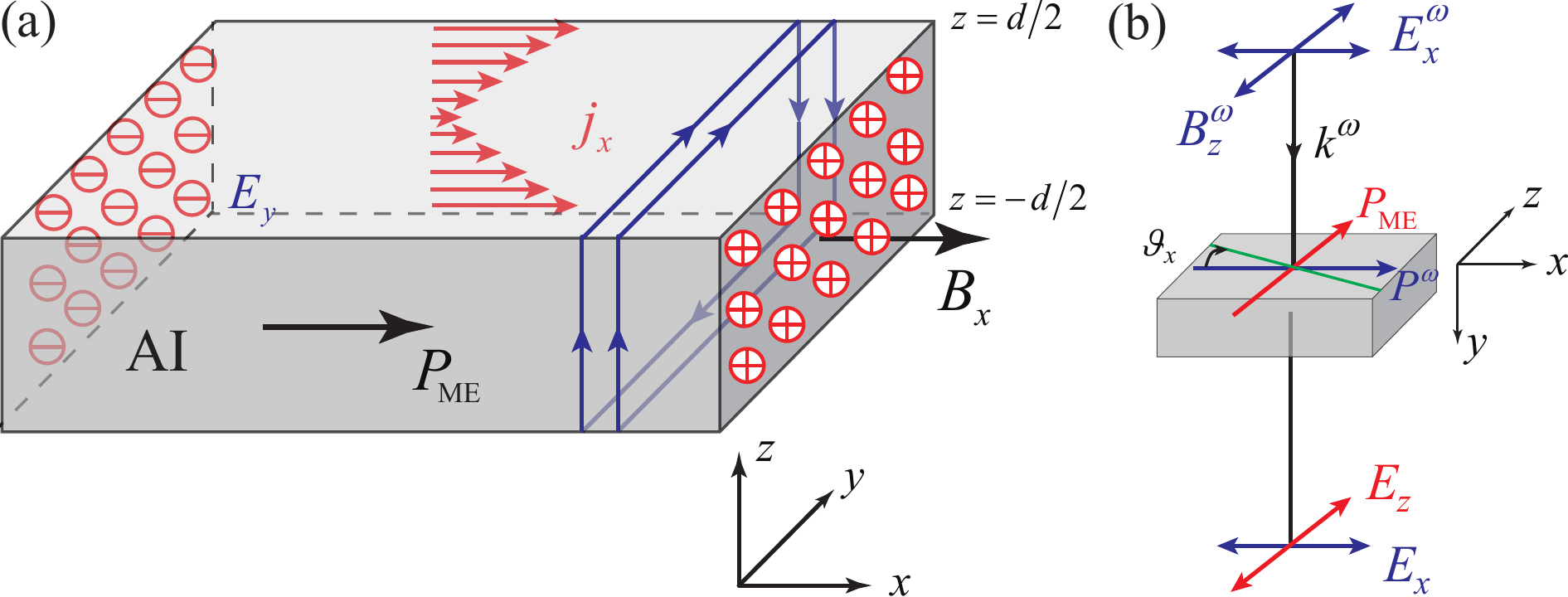}
\end{center}
\caption{(Color online) (a) Illustration of the TME effect, where a parallel charge polarization is induced by a magnetic field. ``$\ominus$'' and ``$\oplus$'' indicate the negative and positive charges induced by the magnetic field on the left and right surfaces, respectively. (b) Schematic of GB with light propagation along $y$ axis. $P_{\omega}$ and $P_{\text{ME}}$ are charge polarization induced by the electric field $E_{\omega}$ and magnetic field $B_{\omega}$ of light, respectively. $\vartheta_x$ is the rotation angle away from principle $x$ axis (denoted as green line).}
\label{fig1}
\end{figure}

The magnetic order of the even SL film belongs to the magnetic point group $3m'$, which allows the diagonal ME susceptibility as follows~\cite{newnham2005}:
\begin{equation}\label{TME}
\hat{\alpha}(d)=\begin{pmatrix}
\alpha_{xx} & 0 & 0\\
0 & \alpha_{xx} & 0\\
0 & 0 & \alpha_{zz}
\end{pmatrix}.
\end{equation}
Here $z$ is defined as the trigonal axis, and $\alpha_{yy}=\alpha_{xx}$. From the symmetry analysis, $\alpha_{xx}\neq\alpha_{zz}$ for finite $d$ in general. As we will show below, only in the 3D bulk limit, $\hat{\alpha}(d\rightarrow\infty)$ becomes isotropic and quantized.

The diagonal TME in Eq.~(\ref{TME}) indicates the induction of a charge polarization when a dc magnetic field is applied,
\begin{equation}\label{polarization}
P_i=-\alpha_{ii}B_i.
\end{equation}
Furthermore, when an ac magnetic field is applied, a parallel polarization current density is induced $j_i=\partial_tP_i$, namely
\begin{equation}\label{current}
j_i=-\alpha_{ii}\partial_tB_i.
\end{equation}

\begin{figure*}[t]
\begin{center}
\includegraphics[width=6.0in,clip=true]{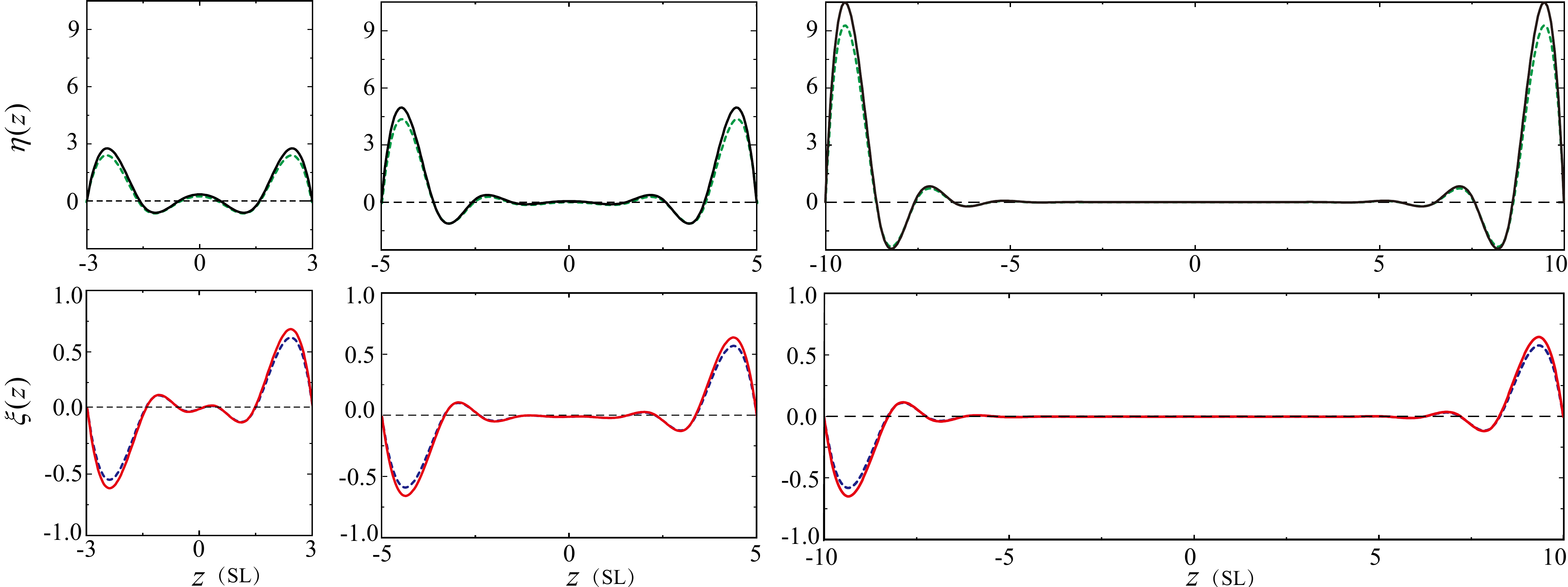}
\end{center}
\caption{(Color online) The TME response functions $\eta(z)$ ($\propto j^{\text{3D}}(z)$) and $\xi(z)$ ($\propto\rho(z)$) for different thickness $6$, $10$, and $20$~SL. The solid and dashed lines denote AI with $m_5=0$ (quantized bulk $\theta$ value) and $m_5\neq0$ (non-quantized bulk $\theta$ value), respectively. The color of lines are consistent with those in Fig.~\ref{fig3}.}
\label{fig2}
\end{figure*}

\emph{Finite-size effect.} The quantized TME described by $\mathcal{L}_{\theta}$ can be understood in terms of a surface Dirac fermion picture~\cite{wang2015b}. Considering the process of applying a magnetic field $\mathbf{B}=B\hat{\mathbf{x}}$ as shown in Fig.~\ref{fig1}(a). A circulating electric field parallel to the side surface due to the Faraday law is generated as $\mathbf{E}^t=-\mathbf{E}^b=\partial_tB(d/2)\hat{\mathbf{y}}$, where the superscript $t$ and $b$ represents top and bottom surfaces, respectively. This then induces a total Hall current density $\mathbf{j}=\sum_{i=t,b}\sigma_{xy}^i\hat{\mathbf{z}}\times\mathbf{E}^i/d$. The surface massive Dirac fermion has half-integer Hall conductance $\sigma_{xy}^t=-\sigma_{xy}^b=(\theta/2\pi)(e^2/h)$. Consequently, a charge density with polarization $\partial_t\mathbf{P}=\mathbf{j}$ is accumulated on the left and right surfaces, namely, a topological contribution to charge polarization $\mathbf{P}=-(\theta/2\pi)(e^2/h)\mathbf{B}$.

Due to the finite-size confinement along the $z$ direction, the surface Dirac fermion on top and bottom surfaces may couple to each other, which leads to the non quantization of $\hat{\alpha}$. The generic Hamiltonian of a AI thin film can be written as $\mathcal{H}_{\text{2D}}(\mathbf{k})=\int_{-d/2}^{d/2}dz\mathcal{H}_{\text{3D}}(\mathbf{k},z)$. Here $\mathbf{k}=(k_x, k_y)$, and we impose periodic boundary conditions in both $x$ and $y$ directions. The ME response $\alpha_{xx}$ of such a thin film along $x$ direction can be directly calculated with the Kubo formula~\cite{wang2015b}. The dc current correlation function
\begin{eqnarray}\label{kubo}
\Pi_{xy}&&(z,z')=\frac{\hbar^2}{2\pi e^2}\int d^2\mathbf{k}\sum_{n\neq m}f(\epsilon_{n\mathbf{k}})
\\
&&\times2\ \text{Im}\left[\frac{\left\langle u_{n\mathbf{k}}\right|j^{\text{3D}}_x(\mathbf{k},z)\left|u_{m\mathbf{k}}\right\rangle\left\langle u_{m\mathbf{k}}\right|j^{\text{3D}}_y(\mathbf{k},z')\left|u_{n\mathbf{k}}\right\rangle}{(\epsilon_{n\mathbf{k}}-\epsilon_{m\mathbf{k}})^2}\right],
\nonumber
\end{eqnarray}
where $\mathbf{j}^{\text{3D}}(\mathbf{k},z)=(e/\hbar)\partial_\mathbf{k}\mathcal{H}_{\text{3D}}(\mathbf{k},z)$ is the 3D in-plane current density operator, $|u_{n\mathbf{k}}\rangle$ is the normalized Bloch wavefunction in the $n$-th electron subband satisfying $\mathcal{H}_{\text{2D}}(\mathbf{k})\left|u_{n\mathbf{k}}\right\rangle=\epsilon_{n\mathbf{k}}|u_{n\mathbf{k}}\rangle$ , and $f(\epsilon)$ is the Fermi-Dirac distribution function. The current density $j^{\text{3D}}_x$ induced by a uniform external ac magnetic field $B_x$ of frequency $\omega/2\pi$ is given by
\begin{equation}\label{3Dcurrent}
j^{\text{3D}}_x(z)=-i\omega\frac{e^2}{2h}\eta(z) B_x\ ,
\end{equation}
where $\eta(z)=2\int_{-d/2}^{d/2}dz_1\ z_1\Pi_{xy}(z,z_1)$ is a dimensionless function. The total 2D current density induced by external magnetic field $B_x$ is given by $j_x^{\text{2D}}=\int dz j^{\text{3D}}_x(z)=-i\omega\gamma_{xx} d(e^2/2h)B_x$, where
\begin{equation}
\gamma_{xx}\equiv\frac{1}{d}\int_{-d/2}^{d/2} dz\ \eta(z),
\end{equation}
Compared to Eq.~(\ref{current}), we get $\alpha_{xx}=\gamma_{xx}(e^2/2h)$. 

Then we calculate $\alpha_{zz}$. The Kubo formula Eq.~(\ref{kubo}) is inapplicable for the vector potential is infinite when the magnetic field is along $z$ axis. The system $\mathcal{H}_{\text{2D}}(\mathbf{k}-e\mathbf{A})=\int_{-d/2}^{d/2}dz\mathcal{H}_{\text{3D}}(\mathbf{k}-e\mathbf{A},z)$ now forms Landau levels (LL), where $\mathbf{A}=B_z(0,x,0)$ is the vector potential in the Landau gauge. The wavefunction for the continuous model $\mathcal{H}_{\text{3D}}$ is written as $\Psi(x,y,z)=\sum_{n,k_y}\varphi_{n}(z)\psi_{n,k_y}(x,y)$ with eigenenergy $\epsilon$, where 
$\psi_{n,k_y}(x,y)$ is 2D LL wavefunction, $n$ is the LL index. Since each LL has the same degeneracy $eB_z/h$, the renormalized charge densities with zero mean along $z$ direction is
\begin{eqnarray}
\rho(z)&\equiv&-\xi(z)\frac{e^2}{h}B_z
\\
&=&-\sum\limits_{\epsilon\leq\epsilon_F}\sum\limits_{n_{occ}}\left(|\varphi_n(z)|^2-\frac{1}{d}\int^{d/2}_{-d/2}|\varphi_n(z)|^2\right)\frac{e^2}{h}B_z,
\nonumber
\end{eqnarray}
where the first summation denotes the sum of eigenstates below the Fermi level, which is fixed at $\epsilon_F=0$. The second summation denotes the sum of occupied LL. $\xi(z)$ is dimensionless. The LL can be obtained by substituting $(k_+,k_-)\rightarrow\sqrt{2}\ell_B^{-1}(a,a^{\dag})$ with the LL lowering and raising operators $a,a^\dag$, where $k_{\pm}=k_x\pm ik_y$, $\ell_B=\sqrt{\hbar/eB_z}$ is the magnetic length. We then diagonalize the Hamiltonian with a LL number cutoff $|a^\dag a|\leq N$, and find $\rho(z)$ quickly converges when $N\geq10$. There are spurious levels due to LL number cutoff $N$, and should not be counted in $n_{occ}$. Thus the charge polarization is $P_z=(1/d)\int dz\ \rho(z)z\equiv-\gamma_{zz}(e^2/2h)B_z$, where
\begin{equation}
\gamma_{zz}=\frac{1}{d}\int^{d/2}_{-d/2}dz\ 2\xi(z)z.
\end{equation}
Compared to Eq.~(\ref{polarization}), we get $\alpha_{zz}=\gamma_{zz}(e^2/2h)$.

The response formulas above are generic for any AI system and do not rely on a specific model. For concreteness, we adopt the effective Hamiltonian in Ref.~\cite{zhang2019} to describe the low-energy bands of MnBi$_2$Te$_4$,
$\mathcal{H}_{\mathrm{3D}}(\mathbf{k},z)=\varepsilon 1\otimes1+d^1\tau_1\otimes\sigma_2-d^2\tau_1\otimes\sigma_1+d^3\tau_3\otimes1-\Delta(z)1\otimes\sigma_3-iA_1\partial_z\tau_1\otimes\sigma_3$. Here $\tau_j$ and $\sigma_j$ ($j=1,2,3$) are Pauli matrices, $\varepsilon(\mathbf{k},z)=-D_1\partial_z^2+D_2(k_x^2+k_y^2)$, $d^{1,2,3}(\mathbf{k},z)=(A_2k_x, A_2k_y, B_0-B_1\partial_z^2+B_2(k_x^2+k_y^2))$, and $\Delta(z)$ is the $z$-dependent exchange field. We then discretize it into a tight-binding model along $z$-axis between neighboring bi-SL from $\mathcal{H}_{\text{3D}}$, and assume $\Delta(z)$ takes the values $\pm\Delta_s$ in the top and bottom layers, respectively, and zero elsewhere. Fig.~\ref{fig2} shows the numerical calculations of $\eta(z)$ and $\xi(z)$ for thin films of $6$, $10$ and $20$~SL. All parameters are taken from Ref.~\cite{zhang2019} for AFM MnBi$_2$Te$_4$, where the surface exchange field $\Delta_s=50$~meV.  Both $\eta(z)$ and $\xi(z)$ are bounded within a finite penetration depth to the top and bottom surfaces. The shapes of the functions $\eta(z)$ and $\xi(z)$ near the surfaces remain almost unchanged as the thickness $d$ varies, which explicitly demonstrates that TME response is from the massive Dirac surface states, and the hybridization between the top and bottom surface states will further deviate TME from quantization. Therefore, TME vanishes for thin films of trivial insulating states (bulk $\theta=0$) without topological surface states.

\begin{figure*}[t]
\begin{center}
\includegraphics[width=6.0in,clip=true]{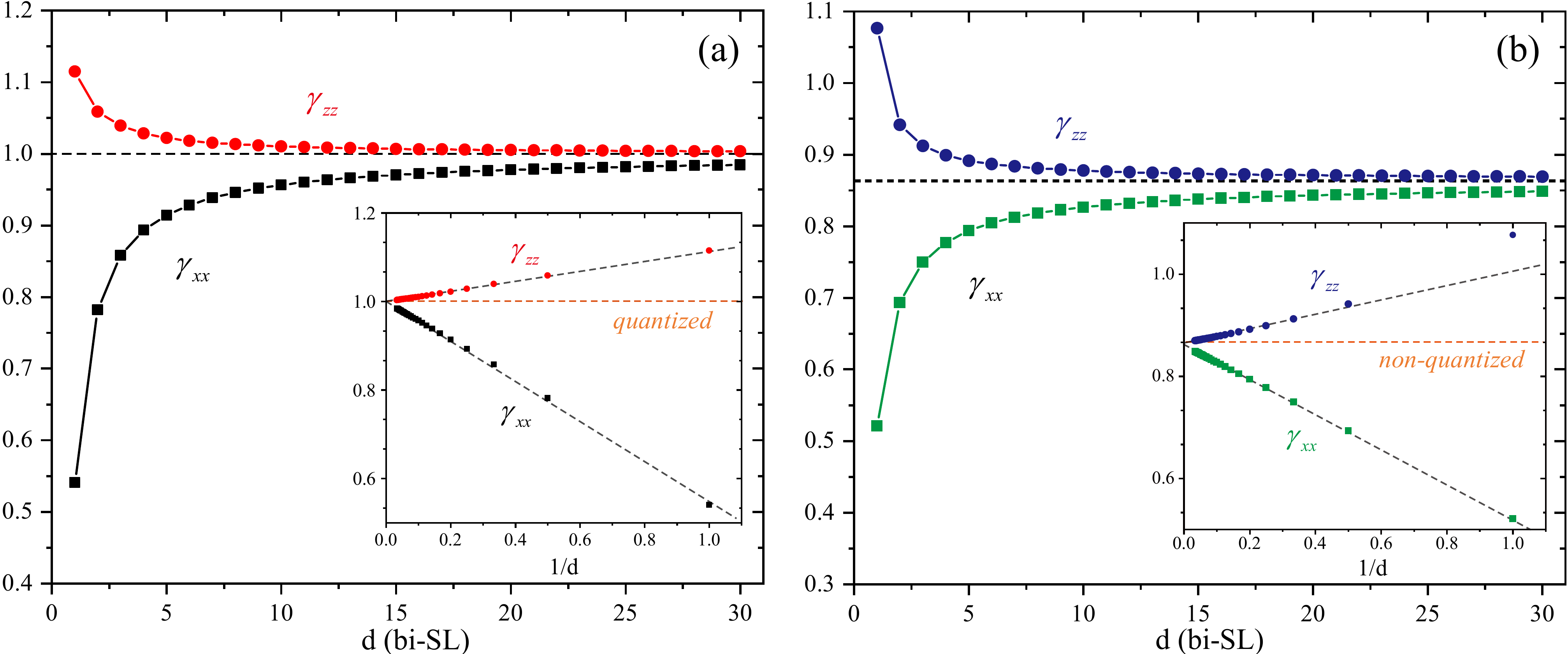}
\end{center}
\caption{(Color online) Finite-size effect of TME effect in the AI thin film with (a) $m_5=0$, and (b) $m_5\neq0$. The $\gamma_{xx}$ and $\gamma_{zz}$ as a function of $d$ show anisotropic behavior. The inset shows $\gamma_{xx}$ and $\gamma_{zz}$ plotted vs the inverse of thickness $1/d$. Here $\theta/\pi=\gamma(d\rightarrow\infty)$.}
\label{fig3}
\end{figure*}

The dimensionless numbers $\gamma_{xx}$ and $\gamma_{zz}$ characterizes the deviation from topological quantization of TME in AI films. The value of $\gamma_{xx}$ and $\gamma_{zz}$ as a function of $d$ is shown in Fig.~\ref{fig3}(a), which shows anisotropic behavior for finite $d$, and both $\gamma_{xx}\rightarrow1$ and $\gamma_{zz}\rightarrow1$ with $d\rightarrow\infty$. This shows that TME effect is quantized as the system is in the thermodynamic bulk limit, consistent with the topological field theory. In fact, as shown in the inset of Fig.~\ref{fig3}(a), the value of $1-\gamma_{xx}$ and $1-\gamma_{zz}$ scales linearly with $1/d$ as the thickness $d\rightarrow\infty$, namely, $1-\gamma_{xx}=\beta_1/d$ and $1-\gamma_{zz}=\beta_2/d$, but with different coefficients.

We emphasize that the anisotropy of TME susceptibility and the $(1-\gamma)\propto1/d$ scaling in finite systems are from the anisotropic shape of thin film, while the scaling coefficient $\beta$ is model parameter dependent. Even we adopt an isotropic Hamiltonian, we still get $\beta_1\neq\beta_2$, since there is no symmetry forbidding it. Furthermore, by adding a $\mathcal{T},\mathcal{P}$-breaking $m_5\tau_2$ term into $\mathcal{H}_{\text{3D}}$~\cite{li2010}, we get an AI state with a non-quantized bulk $\theta$ value such as Mn$_2$Bi$_2$Te$_5$~\cite{wang2016a,zhangj2019}. A direct consequence of the $m_5\tau_2$ term is to open a gap of $2m_5$ in the surface-state spectrum, independent of the surface orientation.  This state is quite different from AI MnBi$_2$Te$_4$ in the bulk. While the former breaks $\mathcal{T,P}$ and bulk $\theta\neq\pi$, while the latter has $\mathcal{S}$ and bulk $\theta=\pi$. Although these two states in thin film have similar transport signature. The dashed lines in Fig.~\ref{fig2} show finite-size TME response with $m_5=20$~meV. The shape of both $\eta(z)$ and $\xi(z)$ are similar to that without $m_5$, but with reduced values. The corresponding $\gamma$ converges to a non-quantized bulk $\theta$ value with $1/d$ scaling as shown in Fig.~\ref{fig3}(b). Here the terminology of TME response for the non-quantized bulk $\theta$ simply means the orbital ME polarizability, it is not related to $Z_2$ topological invariant as in the quantized case $\theta=\pi$. 

\emph{Experimental proposal.}  The TME coefficients in principle can be measured electrically through the polarization current induced by an ac magnetic field. However, such signal is quite small, which is estimated about $0.1$~nA for a sample area of $0.1$~mm$^{2}$ and $B=0.01$~T with $\omega/2\pi=100$~Hz.

Interestingly, the diagonal TME coupling in AI would induce the optical phenomena known as GB, which has been applied to measure the axion-type coupling in multiferroics~\cite{kurumaji2017}. The schematic of GB is illustrated in Fig.~\ref{fig1}(b). The incident light with frequency $\omega/2\pi$ is propagating along $\hat{y}$, the linearly polarized $E^\omega_{\text{in}}//\hat{x}$ induces an oscillating polarization $P^\omega=\chi^e_{xx}E^\omega_z$ along the $x$ axis, where $\chi^e_{xx}$ is the electric susceptibility. While $B^\omega_{\text{in}}//\hat{z}$ induces $P^\omega_{\text{ME}}=\alpha_{zz}B^\omega_{\text{in}}$ along the $z$ axis. The combined $P$ further tilts the electromagnetic wave eigenmodes in the crystal, resulting in a rotation of principal optical axes (the fast and slow axes) around propagation $k^\omega$ direction. The rotation angle $\vartheta_{x}$ ($\vartheta_{z}$) away from the $x$ ($z$) axis is $\tan\vartheta_x=\lambda_1\alpha_{\text{GB}}(\omega)$ ($\tan\vartheta_z=\lambda_2\alpha_{\text{GB}}(\omega)$), where $\alpha_{\text{GB}}(\omega)=\alpha_{xx}(\omega)-\alpha_{zz}(\omega)$, $\alpha_{ii}(\omega)$ is the ME coefficient in the optical frequency range, and parameter $\lambda_{1,2}$ depends on dielectric constant and permeability of the materials~\cite{supplementary}. The electric quadrupole contribution to GB in Fig.~\ref{fig1}(b) configuration vanishes due to crystal symmetry in MnBi$_2$Te$_4$. For low frequency photon with $\omega\ll E_g/\hbar$ ($E_g$ is surface gap), we expect $\alpha_{\text{GB}}(\omega)\approx\alpha_{\text{GB}}(\omega=0)$ (here $\alpha_{ii}(\omega=0)\equiv\alpha_{ii}$). Therefore, the $1/d$ scaling of $\alpha_{\text{GB}}(\omega)$ is an interesting \emph{unique} signature of TME effect. 

For a typical value of $E_{g}=20$~meV, we get $\omega/2\pi\ll4.8$~THz. We further consider the configuration where the AI film is on a non-ME substrate, and still $\vartheta$ of GB is proportional to $\alpha_{\text{GB}}$ with an overall renormalization factor~\cite{supplementary}. The estimation of $\vartheta$ is about $2.1$~mrad for 1~THz light with a sample size of $10$~$\mu$m$\times10$~$\mu$m, which is accessible by the current technique. The side surface states in thin films are gapped due to quantum confinement fulfilling the requirement of TME, which is estimated to be $2A_1\pi/d=0.28/d$~eV. Here $A_1\sim0.6$~eV$\cdot$\AA\ is the velocity along $z$ axis~\cite{zhang2019}, and $d$ is in the unit of number of SLs. Therefore, for thin films of $20$~SLs, it approximately gives a band gap of about $14$~meV on side surfaces. Experimentally, the main difficulty of measuring such an effect is to shining light perpendicular to the film growth direction. While for the incident light along $z$ axis, no GB exists for MnBi$_2$Te$_4$ since $\alpha_{\text{GB}}=\alpha_{xx}-\alpha_{yy}=0$. This may be resolved by searching AI materials with magnetic point group symmetry $222$, $m'm'2$ and $m'm'm'$~\cite{newnham2005}, and the nonzero diagonal TME coefficients $\alpha_{xx}\neq\alpha_{yy}\neq\alpha_{zz}$ for finite $d$, which is left for future work. 

\begin{acknowledgments}
We acknowledge Biao Lian, Yuanbo Zhang, Pu Yu, Shiwei Wu and Yoshinori Tokura for valuable discussions. This work is supported by the Natural Science Foundation of China through Grant No.~11774065, the National Key Research Program of China under Grant Nos.~2016YFA0300703 and 2019YFA0308404, Shanghai Municipal Science and Technology Major Project under Grant No.~2019SHZDZX04, the Natural Science Foundation of Shanghai under Grant No.~19ZR1471400.
\end{acknowledgments}

\end{document}